 \documentclass[manuscript]{aastex}
\usepackage{natbib}
\usepackage{amsmath}

\newcommand{\be}{\begin{displaymath}}
\newcommand{\ee}{\end{displaymath}}

\def\lsim{\hbox{\rlap{\raise 0.425ex\hbox{$<$}}\lower 0.65ex\hbox{$\sim$}}}
\def\gsim{\hbox{\rlap{\raise 0.425ex\hbox{$>$}}\lower 0.65ex\hbox{$\sim$}}}

\shorttitle{Constraints on ICS model}
\shortauthors{Lv et al.}

\begin{document}

\title{Constraint on parameters of the Inverse Compton Scattering model for radio pulsars}

\def\cfa{1}
\def\berk{2}
\def\clay{3}
\def\chikav{4}
\def\chiast{5}
\def\cape{6}
\def\saao{7}
\def\fermi{8}
\def\nd{9}
\def\rut{10}
\def\tok{11}
\def\port{12}
\def\jhu{13}
\def\stsci{14}
\def\penn{15}
\def\psu{16}
\def\stock{17}
\author{M. Lv\altaffilmark{1},  H. G. Wang\altaffilmark{1}, K. J. Lee\altaffilmark{2}, G. J. Qiao\altaffilmark{3}, R. X. Xu \altaffilmark{3}}
\altaffiltext{1}{Center for Astrophysics, Guangzhou University,
Guangzhou 510006, China.}
\altaffiltext{2}{Max-Planck Institute for Radio Astronomy, Bonn 53121, Germany.}
\altaffiltext{3}{Department of Astronomy, Peking University, Beijing 100871, China.}

\email{cosmic008@yahoo.com.cn}

\begin{abstract} The inverse Compton scattering (ICS) model can explain
various pulse profile shapes and diversity of pulse profile evolution
based on the mechanism that the radio emission is generated through
inverse Compton scattering between secondary relativistic particles
and radio waves from polar gap avalanches. In this paper, we study the
parameter space of ICS model for 15 pulsars, which share the common pulse
profile evolution phenomena that the pulse profiles are narrower at higher
observing frequencies. Two key parameters, the initial Lorentz factor and
the energy loss factor of secondary particles are constrained using the
least square fitting method, where we fit the theoretical
curve of the ``beam-frequency mapping'' of the ICS model to the observed pulse
widths at multiple frequencies. The uncertainty of the inclination and viewing angles
are taken into account in the fitting process. It is
found that the initial Lorentz factor is larger than 4000, and the energy
loss factor is between 20 and 560. The Lorentz factor is consistent with the
prediction of the inner vacuum gap model. Such high energy loss
factors suggest significant energy loss for secondary particles at altitudes of
a few tens to hundreds of kilometers.  \end{abstract}


\section{Introduction}

Since the discovery of pulsars, a wealth of observational data have been
accumulated for radio pulsars and the morphological characteristics
of pulsar profile have been widely investigated. The core-double-cone
model (Rankin 1983a,b), a widely used empirical model, attributes the
various kinds of morphologies to geometrical origins. Assuming that the
emission beam consists of a hollow core component, a nesting inner cone,
and an outer cone, the observed single, double, triple, quadruple, and
five-component pulse profiles can be explained as geometrical effects
that the line of sight sweeps across the emission beam from different
locations. Despite the success of the empirical model, physical
mechanisms of radio emission, however, remain open. Among the models of
coherent curvature radiation (Ruderman \& Sutherland 1975, hereafter RS75,
Gil \& Snakowski 1990), plasma instabilities (Asseo et al. 1990, Luo \&
Melrose 1995, Weatherall 1998, Gedalin et al. 2002, Melrose \& Luo 2004),
and inverse Compton scattering (hereafter ICS, Qiao \& Lin 1998, Qiao et
al. 2001), the ICS model manifests itself in simplicity to generate the
frequency-dependent beam structures and flexibility
to reproduce various kinds of frequency dependence of the profile shape
and the pulse width. Beside pulse morphologies, the high brightness
temperature and polarization properties of pulsars can be also explained
by the ICS model (Qiao \& Lin 1998, Zhang et al. 1999, Xu et al. 2000).

In the ICS model, radio emission is generated by the inverse Compton
scattering process between secondary relativistic electrons/positrons
and initial low frequency waves (with frequency $\nu_0\sim 10^{5-6}$
Hz), which are produced by avalanches of the inner vacuum gap. The relation
between the altitude and the Lorentz factor of secondary particles determines
the beam pattern. For example, if the Lorentz factor of secondary
particles decreases as they flow out along open field lines, the emission
at a given frequency would emerge at three different altitudes, which
corresponding to the hollow core, inner and outer cone components. If
the Lorentz factor keeps constant, it would form a beam pattern of only
one hollow core and one cone (the inner cone). The ICS model predicts that,
in the outer cone, the emission with a higher frequency is generated at
a relatively lower altitude, while the radiation in the inner cone follows
the opposite relation. Observationally, the anti-correlation between the pulse
width and the frequency is found in many pulsars with conal components, although
a small amount of pulsars show constant or increasing pulse widths
at higher frequencies. The ICS model agrees with these observations,
and attributes these two types of relation to the outer and inner conal components,
respectively (Qiao et al. 2001).

 In order to constrain the parameter space of ICS model, some authors
have already compared the predictions of the ICS model with the observational
data. There are three basic parameters for the ICS model, the initial Lorentz
factor $\gamma_0$, the frequency of initial radio wave $\nu_0$, and the
energy loss factor $\xi$ for the secondary particles. Lee et al. (2009,
hereafter LCW09) found that the ratio between the radiation altitudes at different altitudes is
insensitive to inclination angles $\alpha$ for radio pulsars with large
linear polarization position-angle (PA) swing rate.  Assuming that the
Lorentz factor decreases with altitude, the authors constrained the parameter
space of $\gamma_{0}$ and the energy loss factor $\xi$ for five radio
pulsars by fitting the ratios of profile width at multi frequencies. Four
out of five pulsars show clear decreasing pulse width with frequency,
while the other one shows nearly constant pulse width.

In this paper, we extend the above reverse-engineering test to a larger sample of
pulsars and check the parameter space of the ICS model, where 15 pulsars with
anti-correlations between the pulse width and the frequency is selected to check the
radiation behavior of the conal beam in the ICS model. We calculate the
``beam-frequency mapping'' of ICS model in Section 2. The geometrical method to
calculate the beam width from the pulse width is presented in
Section 3. Details about data reduction and related results are given in
Section 4. Conclusions and discussions are summarized in Section 5.

\section{The beam-frequency mapping in the ICS model}

In the ICS model, low-frequency waves $\nu_{0}$ are excited by the
periodic breakdown of the inner vacuum gap. The waves are then inversely
Compton scattered by the relativistic secondary particles produced
in the pair cascade process in the gap. The scattered waves have the frequency:
\begin{equation} \nu=2\gamma^{2}\nu_{0}(1-v\cos\theta_{\rm i}/c)\,,
	\label{eq:icsf}
\end{equation}
where $v$ is the velocity of the particle and $\theta_{\rm i}$ is the
incident angle between the particle motion direction and the incoming photons, and
$\nu_0$ is assumed to be $10^{6}$ Hz. There are a few physical
considerations for the value of $\nu_0$. RS75 pointed out that the growth
and the fluctuation time scale of the inner vacuum gap might be (30-40)$\tau_0$,
where $\tau_0$ is the time needed for conversion of a gamma-ray photon
to a pair in the gap, which has the order $\sim$(gap height)$/c$. The
fluctuation time scale is estimated to be about 10 microseconds. But
it is not conclusive yet even with the recent simulation (Timokhin
2010). Observationally, durations of a few microsecond was indeed
observed in some pulsars (e.g. Bartel \& Sieber 1978, Lange et al. 1998).

For the geometrical configuration in
Figure~1, when the radiation altitudes are far from the pulsar surface, we have
(Qiao et al. 2001) \begin{equation} \cos\theta_{\rm
	 i}=\frac{2\cos\theta-(R/r)(1-3\cos^{2}\theta)}{\sqrt{(1+3\cos^{2}\theta)[1-2(R/r)\cos\theta+(R/r)^{2}]}}\,,
\end{equation} where $r$ is the distance between the scattering point and the
center of the pulsar, $R$ is pulsar radius, $\theta$ is the polar angle between
the radiation location and the magnetic axis. For a dipole magnetic field, we have \begin{equation} r=R_{\rm e}\sin^{2}\theta\,,
\end{equation} where $R_{\rm e}$ is the maximal radius of a given magnetic field
line. The angular beam width $\theta_\mu$ of the radiation beam coming from the
place with polar angle $\theta$ is,
\begin{equation}
	\tan\theta_{\mu}=\frac{3\sin2\theta}{1+3\cos2\theta},
\end{equation}
where $\theta_{\mu}$ is the angle between the direction of magnetic field at the point Q
and the magnetic axis (Figure~1). From the radiation altitude $r$ and Lorentz
factor $\gamma$, one can determine the outgoing radio wave frequency $\nu$ and
the angular beam width $\theta_{\mu}$ using the above equations.
	
The energy of the secondary particles decrease as they flow along the field lines,
it is usually assumed that the Lorentz factor follows \begin{equation}
	\gamma=\gamma_{0}[1-\xi(r-R)/R_{\rm e}]\,, \label{eq:gamma_ics}
\end{equation} where $\gamma_{0}$ is the initial Lorentz factor at the top of
the gap and $\xi$ is the energy loss factor. From Equations~(\ref{eq:icsf}) and
(\ref{eq:gamma_ics}), one can see that
$\nu_0$ and $\gamma_{0}$ are degenerated, i.e. different groups of $\nu_0$
and $\gamma_0$ can give identical results in the ICS model. Thus, in the
subsequent test for the ICS model, we fix the value of $\nu_0$ without losing generality.

Using the above equations, the relation between the outgoing frequency $\nu$ and
the beam width $\theta_{\mu}$, i.e.  the function
$\nu=\nu(\nu_{0},\gamma_{0},\xi,\theta_{\mu})$, can be calculated. We refer this
$\nu$-$\theta_{\mu}$ relation as the ``beam-frequency mapping''. A characteristic beam-frequency mapping with the dropping-rising-dropping pattern is given in
Figure~2. Compared with the empirical core-double-cone framework for pulsar
radiation, the leftmost branch corresponds to the core component, the middle
rising part corresponds to the inner cone and the rightmost part corresponds to
the outer cone. In this paper, we fit such beam-frequency mapping to the
observational data to infer the parameters of ICS model, where the fitting
method is given in next section.

\section{Method}

 In our method, we first determine $\theta_{\mu}$ from the pulse width using
conventional geometrical model \citep{gil84, lyn88}, and then
fit the beam-frequency mapping to $\theta_\mu$ data at multiple frequencies to infer the
parameters. Here the pulse width $\phi$ is measured using the Gaussian
decomposition method (Wu et al. 1992, Kramer et al. 1994a,b, LCW09).

\subsection{Measurement of pulse width}
 The Gaussian decomposition method is widely used to measure the pulse width of
a pulse profile, where one models the profile with multiple Gaussian
components and extracts the pulse width using curve fitting techniques, i.e. one
fits the pulse profile with the following form \citep{wu92, kra94a, kra94b}
\begin{equation} I=\sum_{i=1}^{n} I_{i}{\,\rm
	exp}{\left[-\frac{(\phi-\phi_{i})^{2}}{2\sigma_{i}^{2}}\right]}\,.
\end{equation} Here we denote the intensity of pulse profile $I(\phi)$ as a function of
longitudinal phase $\phi$, which is modeled with $n$ Gaussians. The $I_{i}$ is
the amplitude of the $i$-th Gaussian, $\phi_{i}$ and $\sigma_{i}$ are the
central phase and standard deviation of Gaussian component respectively.

We use the 10\% width throughout this paper, which is
defined as the full pulse width $2\Delta\phi$ between the leading phase
of the leftmost component and the trailing phase of the rightmost component down
to the 10\% level of their maximal intensities.

The number of Gaussians, the amplitude, the peak phase and the typical
width of each Gaussian are free parameters. Following LCW09, we use
Levenberg-Marquardt method to perform the least square fitting, which is
accepted only when i) nonparametric Kolmogorov-Smirnov test is passed when
comparing with the distributions of residues in the on-pulse and off-pulse
regions, and ii) the difference between the rms levels of two residues is
close to zero, i.e.  $\eta=|\sigma_{\rm on}-\sigma_{\rm off}|/\sigma_{\rm
off}\simeq 0$. The number of Gaussians is determined by minimizing
$\eta$. We run a small Monte-Carlo simulation to determine the final fitting
parameters, where the fitting is repeated several times (about 10) with randomly generated
initial values, and the final parameters are the averaged values from each
individual accepted fitting. We list the measured final pulse width in Table 1.

\subsection{Determining angular sizes of radiation beams}

From the pulse width, we calculate the angular size of radiation
beam following the conventional geometrical model, which assumes that, i) the
magnetic field in the magnetosphere is dipolar, and ii) the radiation direction is
parallel to the local
magnetic field (Figure~3). When the line of sight locates in the
$\Omega-\mu$ plane, the rate of linear polarization PA
swing reaches the maximum $\kappa$ \citep{rad69}, where \begin{equation}
\kappa=\frac{\sin\alpha}{\sin\beta}\,. \label{eq:pa} \end{equation}
Here $\alpha$ is the inclination angle, the impact angle $\beta$
is defined as the angle between the line of sight and the magnetic axis.

Given the inclination angle $\alpha$ and $\kappa$, we can determine $\beta$. We
can further substitute $\beta$, $\Delta\phi$ and $\alpha$ into the
following geometrical relation to calculate the beam width $\theta_{\rm \mu}$
\citep{gil84, lyn88}
\begin{equation}
\sin^{2}\left(\frac{\theta_{\mu}}{2}\right)=\sin^{2}\left(\frac{\triangle\phi}{2}\right)\sin\alpha
	\sin(\alpha+\beta)+\sin^{2}\left(\frac{\beta}{2}\right)\,.
	\label{eq:beamangle}
\end{equation}

\subsection{Measurement for ICS model parameter}

 The ICS model parameters are derived via fitting the $\theta_\mu$ data at multiple frequencies
with the beam-frequency mapping. On the one hand, given $N$
frequency bands at central frequencies of $\nu_{i}$, we can measure $N$ beam
widths $\theta_{\mu, i}$ from the observation, where $i=1\cdots N$.  On the other
hand, with Equations~(1)-(4), we can calculation the predicted beam width $ \theta_{\mu,{\rm i}}^{\rm
ics}$ of the ICS model at those frequencies as a function of parameters $\gamma_0$ and
$\xi$. Thus we can fit the ICS model to the data by minimizing the following
standard $\chi^2$, \begin{equation}
\chi^2=\sum_{\rm
i=1}^{\rm N}\left(\frac{\theta_{\mu,{\rm i}}-\theta_{\mu,{\rm i}}^{\rm
ics}}{\delta\theta_\mu}\right)^2\,,
\end{equation}
where $\delta\theta_\mu$ is the error of $\theta_\mu$. In our fitting, the
third ICS model parameter, the background wave frequency $\nu_{0}$, is fixed as
$10^{6}$ Hz. Such fixation is due to the parameter degeneracy, i.e. the effect
of $\nu_0$ is completely absorbed into the parameter $\gamma_0$ as indicated in
the Equation~(1).


\section{Data reduction and results}

 The pulse profile data are from the European Pulsar Network Database,
where 15 pulsars are selected according to two criteria, viz. 1) high pulse
profile quality at more than five frequencies ranging over one order
of magnitude, (2) the absolute values of the maximum slope rate of
linear polarization PA $\kappa$ is much larger than 1. The second requirement,
i.e. $\kappa\gg1$, is to make sure that the ratio between angular sizes of
radiation beams at different frequencies are insensitive to the inclination angle \citep{lee09}.

As explained in the previous section, given the inclination angle $\alpha$, the maximal
PA swing slope $\kappa$, and the pulse width, we can calculate the angular sizes
$\theta_{\mu}$ of the radiation beams at multiple frequencies. With the
beam-frequency mapping, we fit for the ICS model parameters.

One caveat is that it is difficulty to measure the inclination angle accurately.
Two types of methods have been used to estimate $\alpha$. The first type of
methods are fitting the polarization PA data with the conventional
\citep{rad69,lyn88, eve01} or modified versions
(Blaskiewicz et al. 1991) of the rotation vector model. The second type of methods are based on some statistical
relations between the pulse period and the opening angle of emission beam
(Rankin 1990, Kijak \& Gil 1997). Due to the limitation of each method
and data quality, in many cases, different authors got inconsistent
values of inclination angle. Two major methods are used here to reduce the effect of uncertainty
of the inclination angle. Firstly, we select pulsars with larger $\kappa$. As
proved by LCW09, although the absolute beam width relies on the inclination
angle, the ratio between the beam widths at multiple frequencies is insensitive
to the inclination angle. Such invariant ratio is already able to determine the
ICS model parameters. Secondly, we perform the measurement of $\theta_{\mu}$
and the fitting with the ICS model for all the possible $\alpha$ values, of which the range
is presented in Table 2. By enumerate all the possible $\alpha$, we
ensure that the invalid parameter space is really not viable.

We show the allowed regions in the $\gamma_0$--$\xi$ parameter space in
Figure~\ref{fig4}, where the contours are at 2-$\sigma$ level. We also show the
result from the method in LCW09. The results immediately show that fitting the
absolute values of
$\theta_\mu$ can tell us how the allowed parameter space depends
on the inclination angle. However, it is not sensitive to constrain
$\gamma_0$, because the outer-cone branches of beam-frequency curves
with the same $\xi$ but different $\gamma_0$ are compressed very much along the
$\theta_\mu$-dimension, as presented in Figure.~\ref{fig2}.  The constrained
parameter ranges are also listed in Table 2. The general features are summarized as
follows.

(1) Combining with all the samples, the allowed parameter space are
$\gamma_0>4000$ and $20<\xi<560$. Since the Lorentz factor of secondary
particles is likely well below $10^5$ according to inner vacuum gap
model, we set the cutoff of $\gamma_0$ to be 20000 in calculation.
No clear correlation is found between the constrained parameter values
and observational quantities, e.g. the pulse width, the surface magnetic field,
the pulsar age and the spin-down energy loss rate.

(2) It shows a general trend that both the initial Lorentz factor
$\gamma_0$ and the energy loss factor $\xi$ increase as $\alpha$
decreases. This is because $\theta_\mu$ and the emission altitude $r$ become
smaller when $\alpha$ decreases, it requires a faster energy loss so that
the Lorentz factor decreases to proper values to produce the radio emission
at the observed frequencies.

\clearpage \section{Conclusions and Discussions}

We collected a sample of 15 pulsars which have the anti-correlation
phenomenon between the pulse width and the observing frequency. Their pulse
widths are measured from multi-frequency profiles by using the Gaussian
decomposition techniques. Beam widths are calculated with the classical
beam geometry model for possible inclination angles. Then they are
fitted with the beam-frequency relation of the ICS model to constrain two
free parameters, i.e. the initial Lorentz factor $\gamma_0$ and the energy loss
factor $\xi$. The fitting is performed for a group of possible $\alpha$. It
shows a trend that $\gamma_0$ and $\xi$ could be larger for smaller
inclination angles. The allowed parameter space are $\gamma_0>4000$ and
$20<\xi<560$. The range of the initial Lorentz factor is generally consistent with
the prediction of the inner vacuum gap model (Timokhin 2010). The
constrained values of $\xi$ suggest that the secondary particles need
to lose a large fraction of their initial energy.

In our calculations, we assume $\nu_0=10^6$ Hz without losing
generality due to the parameter degeneracy, which also agrees with the physical
consideration of the inner vacuum gap model. If we take $\nu_0=10^5$ Hz,
according to Equation~(\ref{eq:icsf}), the resulted Lorentz factors would be
about 3 times higher than the present results.

Our results indicates a bit higher Lorentz factor $\gamma_0>4000$ than the traditional
picture of the inner vacuum gap model ($\sim1000$ in RS75), but this agrees with the requirement to produce the
radio luminosity of pulsars in the ICS model. Noting that the primary
particles usually has a Lorentz factor of $10^{6}\sim10^7$ due to radiation
reaction (e.g. RS75), the present results imply that the multiplicity of
secondary particles is about a few hundred. With such a multiplicity,
only a small fraction of secondary particles being coherent (Qiao \&
Lin 1998) or even incoherent radiation in some cases (Zhang et al. 1999)
is sufficient to account for the observed radio luminosity, because
the ICS emission of a single particle is much more efficient than the
curvature radiation. \cite{HM11} found that a slightly asymmetric
distortion can significantly increase the accelerating electric field on one side of
the polar cap and, combined with a smaller field line radius of curvature, would lead
to larger pair multiplicity. This increase of the primary accelerating
electric filed may also be the origins of high Lorentz factors of the secondary particles.

The result of $20<\xi<560$ indicates an efficient energy lose for the
secondaries. It has been suggested that the resonant inverse Compton
scattering between secondaries and thermal photons from the pulsar surface
can cause efficient energy loss at a certain surface temperature,
but it is only effective within about one stellar radius above the polar
cap surface (Zhang et al. 1997, Lyubarskii \& Petrova 2000). The other
emission mechanisms are much less efficient (Sturner 1995) for secondaries, hence can
be neglected. Therefore, the mechanism for such efficient energy loss is
still unclear. LCW09 revealed that the decay of secondary Lorentz factor
varies significantly from pulsar to pulsar. For some pulsars (e.g. PSR
B2016+28), of which the profile width nearly keeps constant at different frequencies,
a small value of $\xi$ is thus inferred. Such diversity of energy
loss behaviors may due to the interplay between the residual electrical field and the
radiation reaction at higher altitudes, of which detailed studies is still
demanded.

\section{Acknowledgement}
We are grateful to the anonymous referee for his/her valuable comments.
The work is partially supported by the Bureau of Education of Guangzhou
Municipality (No.11 Sui-Jiao-Ke [2009]), GDUPS (2009), Yangcheng
Scholar Funded Scheme (10A027S)), NSFC (10778714, 10833003, 10973002,
10821061, 10935001) and the National Basic Research Program of China
(Grant 2009CB824800). K. J. Lee is also supported by ERC Grant ¡°LEAP¡±,
Grant Agreement Number 227947 (PI Michael Kramer).

\clearpage

\begin{center}
\begin{deluxetable}{llllllll}
\tablecaption{Pulsar parameters and their 10\% pulse widths as functions of observing frequencies}
\tablehead{\colhead{Name} &
\colhead{$P$\,(s)} & \colhead{$B$\,($10^{12}$\,G)}  & \colhead{$\kappa$}    &\colhead{$\nu$\,(Hz)}   &\colhead{2$\Delta\phi$\,($^\circ$)}   &\colhead{$\eta$} &\colhead{Reference} }
\startdata
B0301+19               &1.388 &1.36l                   &-17   &0.408  &20.2$\pm$3.8     &0.206 &  GL98\\
                       &      &                        &      &0.61   &19.6$\pm$1.1     &0.078 &  GL98 \\
                       &      &                        &      &0.925  &18.9$\pm$0.7     &0.189 &  GL98\\
                       &      &                        &      &1.408  &17.7$\pm$0.7     &0.133 &  GL98\\
                       &      &                        &      &1.642  &16.3$\pm$0.5     &0.154 &  GL98 \\
			           &      &                        &      &4.85   &15.8$\pm$0.4     &0.035 &   HX97\\ \\
B0525+21               &3.745 &12.4                    &31    &0.408  &21.7$\pm$1.9     &0.031 &  GL98 \\
                       &      &                        &      &0.61   &22.1$\pm$0.7     &0.303 &  GL98\\
                       &      &                        &      &0.925  &21.7$\pm$2.2     &0.347 &  GL98\\
                       &      &                        &      &1.642  &20.9$\pm$0.7     &0.458 & GL98\\
				   	   &      &                        &      &4.85   &17.7$\pm$0.6     &1.607 & HX97\\ \\
B0540+23               &0.246 &1.97                    &-9    &0.91   &43.2$\pm$2.8     &0.068 & GL98  \\
                       &      &                        &      &1.408  &41.9$\pm$2.0     &0.235 & GL98\\
                       &      &                        &      &1.642  &43.7$\pm$2.9     &0.044 &GL98 \\
					   &      &                        &      &4.85   &38.8$\pm$1.4     &0.069 & HX97\\
					   &      &                        &      &10.45  &34.0$\pm$4.0     &0.094 & HX97\\ \\
B0823+26               &0.531 &0.96                    &-4.5  &0.408  &21.1$\pm$4.3     &0.067 & GL98  \\
                       &      &                        &      &0.61    &21.2$\pm$2.4     &0.369 & GL98\\
                       &      &                        &      &0.925  &18.9$\pm$2.5     &0.397 & GL98 \\
                       &      &                        &      &1.404    &12.8$\pm$0.7     &0.839 & GL98 \\
                       &      &                        &      &4.85    &12.7$\pm$1.2     &0.727 &  HX97\\
                       &      &                        &      &10.55  &10.9$\pm$0.3     &0.803 &  SGG95\\\\
B0919+06               &0.431  &2.45                   &7     &0.408  &27.9$\pm$1.6     &0.290 & GL98 \\
                       &      &                        &      &0.61   &22.8$\pm$1.5     &0.411 & GL98 \\
                       &      &                        &      &0.925  &17.1$\pm$0.6     &0.115 & GL98 \\
                       &      &                        &      &1.408  &18.8$\pm$1.2     &0.070 & GL98 \\
                       &      &                        &      &1.642  &15.6$\pm$2.4     &0.087 & GL98 \\
                       &      &                        &      &4.85   &10.4$\pm$1.1     &0.233 &  HX97\\
                       &      &                        &      &10.55  &7.6 $\pm$0.3     &0.149 &  SGG95\\\\
B1039-19               &1.386 &1.16                    &-18   &0.4    &24.8$\pm$1.3    &0.229 &  ANT94\\
                       &      &                        &      &0.61   &20.3$\pm$0.4     &0.394 &  GL98 \\
					   &      &                        &      &0.8    &22.4$\pm$1.0     &0.139 &  ANT94\\
                       &      &                        &      &0.925  &19.5$\pm$1.0     &0.230 &  GL98\\
                       &      &                        &      &1.408  &17.6$\pm$0.5     &0.147 &  GL98 \\
                       &      &                        &      &1.642  &18.1$\pm$0.8     &0.326 &   GL98\\
                       &      &                        &      &4.85   &17.6$\pm$1.2     &0.032 &  KKW97 \\\\
B1133+16               &1.188 &2.13                    &12    &0.102  &22.9$\pm$1.3    &1.633 & KL99\\
                       &      &                        &      &0.408  &14.7$\pm$0.6     &0.095 & GL98  \\
                       &      &                        &      &0.925  &13.4$\pm$0.7     &0.077 &  GL98  \\
                       &      &                        &      &1.408  &14.0$\pm$1.5     &0.959 & GL98   \\
                       &      &                        &      &1.71   &12.3$\pm$0.8     &0.614 &   HX97  \\
					   &      &                        &      &2.25   &11.1$\pm$0.4     &0.645 &  KXJ96  \\
                       &      &                        &      &4.85   &11.2$\pm$0.6     &2.485 &    HX97\\
                       &      &                        &      &10.45  &9.1$\pm$0.8      &0.577 &   HX97\\\\
B1702-19               &0.299 &1.13                    &-14   &0.61   &20.3$\pm$0.7    &0.388 & GL98\\
                       &      &                        &      &0.925  &18.2$\pm$0.9     &0.121 & GL98 \\
                       &      &                        &      &1.408  &17.4$\pm$0.4     &0.437 & GL98 \\
                       &      &                        &      &1.642  &17.2$\pm$0.4     &0.407 & GL98 \\
                       &      &                        &      &4.85   &16.7$\pm$0.4     &0.091 & KKW97\\\\
B1706-16               &0.653 &2.05                    &-9     &0.408  &14.8$\pm$1.0   &0.272 & GL98   \\
                       &      &                        &       &0.61   &14.2$\pm$0.6    &0.251 & GL98 \\
                       &      &                        &       &0.925  &14.6$\pm$0.8    &0.300 & GL98  \\
                       &      &                        &       &1.408  &14.5$\pm$0.5    &0.065 & GL98  \\
                       &      &                        &       &4.75   &10.8$\pm$0.5    &0.103 &  SGG95\\\\
B2020+28               &0.343  &0.82                    &15     &0.41   &18.5$\pm$0.5   &0.522 &  GL98 \\
                       &      &                        &       &0.925  &18.5$\pm$0.6    &2.365 & GL98 \\
                       &      &                        &       &1.408  &19.7$\pm$0.4    &0.549 & GL98 \\
                       &      &                        &       &1.642  &19.4$\pm$0.6    &1.893 & GL98 \\
                       &      &                        &       &4.85   &18.6$\pm$0.8    &1.780 &  HX97\\
                       &      &                        &       &10.45  &14.9$\pm$0.7    &0.275 &  HX97\\\\
B2021+51               &0.529  &1.29                    &4      &0.41   &25.6$\pm$1.4   &0.489 & GL98   \\
                       &      &                        &       &0.925  &22.3$\pm$0.5    &0.788 & GL98  \\
                       &      &                        &       &1.408  &22.9$\pm$0.8    &1.567 & GL98  \\
                       &      &                        &       &4.75   &17.8$\pm$0.3    &0.177 &   SGG95\\
                       &      &                        &       &10.45  &13.0$\pm$0.4    &0.045 &   HX97\\
					   &      &                        &       &14.6   &12.7$\pm$0.6    &0.027 &   KXJ96\\\\
B2045-16               &1.962 &4.69                    &-30    &0.606  &17.8$\pm$0.2    &2.131 &  GL98 \\
                       &      &                        &       &0.925  &17.9$\pm$0.8    &0.080 &  GL98\\
                       &      &                        &       &1.408  &16.8$\pm$0.3    &1.746 &  GL98\\
                       &      &                        &       &1.642  &16.4$\pm$0.4    &1.788 & GL98 \\
                       &      &                        &       &4.85   &15.7$\pm$0.5    &0.293 & HX97 \\\\
B2154+40               &1.525 &2.32                    &8      &0.408  &32.1$\pm$2.1    &0.410 &  GL98  \\
                       &      &                        &       &0.61   &31.4$\pm$0.7    &0.294 &  GL98 \\
                       &      &                        &       &0.925  &28.8$\pm$1.4    &0.105 &  GL98 \\
                       &      &                        &       &1.408  &29.9$\pm$2.0    &0.108 &  GL98  \\
                       &      &                        &       &1.642  &27.4$\pm$1.2    &0.191 &  GL98 \\
                       &      &                        &       &4.85   &26.3$\pm$1.2    &0.178 &  HX97\\\\
B2310+42               &0.349 &0.20                    &4       &0.408  &19.4$\pm$1.2   &0.049 & GL98 \\
                       &      &                        &        &0.61   &18.4$\pm$0.9   &1.417 & GL98 \\
                       &      &                        &        &0.925  &19.6$\pm$1.1   &0.194 & GL98 \\
                       &      &                        &        &1.408  &17.8$\pm$0.7   &1.581 & GL98 \\
					   &      &                        &        &1.615  &17.6$\pm$0.5   &1.419 &  SGG95 \\
                       &      &                        &        &4.85   &15.7$\pm$0.3   &0.221 &  HX97\\
                       &      &                        &        &10.55  &15.2$\pm$0.8   &0.127 &  SGG95\\\\
B2319+60               &2.256 &4.03                    &-8     &0.925  &24.7$\pm$0.8    &0.079 &  GL98 \\
                       &      &                        &       &1.408  &23.6$\pm$1.0    &0.379 & GL98 \\
                       &      &                        &       &1.642  &22.6$\pm$0.3    &0.188 & GL98 \\
                       &      &                        &       &4.85   &19.8$\pm$0.6    &0.146 &  HX97\\
                       &      &                        &       &10.55  &20.6$\pm$1.4    &0.554 &  SGG95\\
											 \hline
\enddata
\tablerefs{Pulsar parameters of the periods and the surface magnetic field
intensities are taken from ATNF pulsar catalogue \citep{MHTH05}. The maximal slops of PA
curves $\kappa$ are from \cite{lyn88}. The original
references for the profile data are given in the last column of the table as the following abbreviations:
ANT94-Arzoumanian et al. 1994, GL98-Gould \& Lyne 1998, HX97-Hoensbroech \& Xilouris 1997, KKW97-Kijak et al. 1997,
KL99-Kuz'min \& Losovskii 1999, KXJ96-Kramer et al. 1996, SGG95-Seiradakis et al. 1995. } 
\end{deluxetable}
\end{center}

\clearpage

 \begin{deluxetable}{lcccl}
 \tablecolumns{8} \tablewidth{0pc} \tablecaption{Range of the inclination angle $\alpha$ and the constrained joint parameter space.} \tablehead{ \colhead{Name} &
 \colhead{$\alpha$($^\circ$)}  &\colhead{Range of $\xi$}   &\colhead{Range of $\gamma_{0}$}    &\colhead{References for possible $\alpha$.} }
 \startdata
 B0301+19       &5$\sim$90    &50$\sim$420  &5000$\sim$20000     & 1 2 3 5 10      \\
 B0525+21       &14$\sim$67   &50$\sim$560  &5500$\sim$20000     & 1 4 5 6 8 9 10    \\
 B0540+23       &17$\sim$90   &10$\sim$90   &4000$\sim$16000     & 1 2 4 8               \\
 B0823+26       &60$\sim$90   &26$\sim$50   &5800$\sim$10000     &1 2 5 6 9               \\
 B0919+06       &58$\sim$90   &34$\sim$80   &4200$\sim$7600      &1 8           \\
 B1039-19       &23$\sim$31   &175$\sim$410 &11000$\sim$20000    &5 10     \\
 B1133+16       &33$\sim$60   &80$\sim$425  &6500$\sim$20000     & 1 5 6 8 9   \\
 B1702-19       &75           &64$\sim$110  &10000$\sim$20000    & 5    \\
 B1706-16       &40$\sim$70   &70$\sim$290  &6000$\sim$20000     & 5 6    \\
 B2020+28       &60$\sim$71   &60$\sim$130  &9200$\sim$20000     & 5 6    \\
 B2021+51       &16$\sim$30   &68$\sim$170  &11500$\sim$20000    &9     \\
 B2045-16       &24$\sim$90   &70$\sim$375  &8000$\sim$20000     &2 5     \\
 B2154+40       & 15          &210$\sim$430 &12000$\sim$20000    &5       \\
 B2310+42       &24$\sim$56   &25$\sim$175  &4000$\sim$15000     &5 6 7     \\
 B2319+60       &12$\sim$35   &85$\sim$245  &10800$\sim$20000    &5 8     \\
 \hline
 \enddata
\tablerefs{1-Blaskiewicz et al (1991); 2-Hoensbroech $\&$ Xilouris (1997); 3-Narayan $\&$ Vivekanand (1982); 4-Gould (1994); 5-Kuz'min $\&$ Wu (1992); 6-Lyne $\&$ Manchester (1988); 7-Wang $\&$ Wu (2003); 8-Kijak $\&$ Gil (1997);9-Rankin (1993);  10-Everett $\&$ Weisberg (2001) }
 \end{deluxetable}

\clearpage

\begin{figure}
\includegraphics[width=0.5\hsize,height=0.5\hsize]{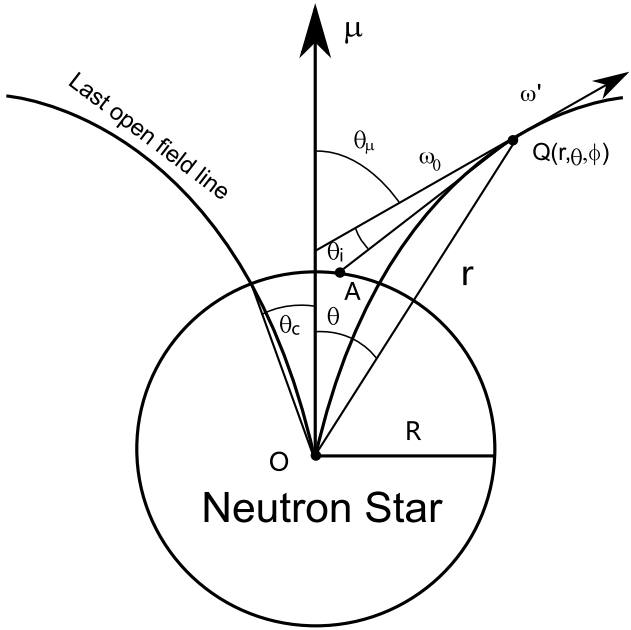}
\caption{Geometry of the inverse Compton scattering model (Qiao \& Lin 1998). \label{fig1}}
\end{figure}

\begin{figure}
\includegraphics[width=0.65\hsize,height=0.5\hsize]{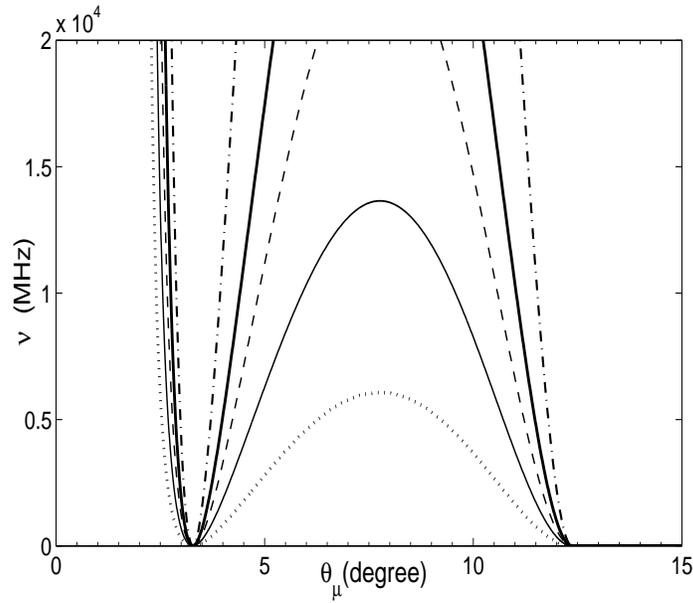}
\caption{The ``beam-frequency mapping'' in the ICS model. For the curves from bottom to top, the corresponding $\gamma_0$ are 4000, 6000, 8000, 10000, 15000, while $\xi$ is fixed at 50. It shows that $\theta_\mu$ at a given frequency in the outer-cone branch becomes insensitive to $\gamma_0$ when $\gamma_0$ is large enough.
\label{fig2}}
\end{figure}

\begin{figure}
\includegraphics[width=0.5\hsize,height=0.5\hsize]{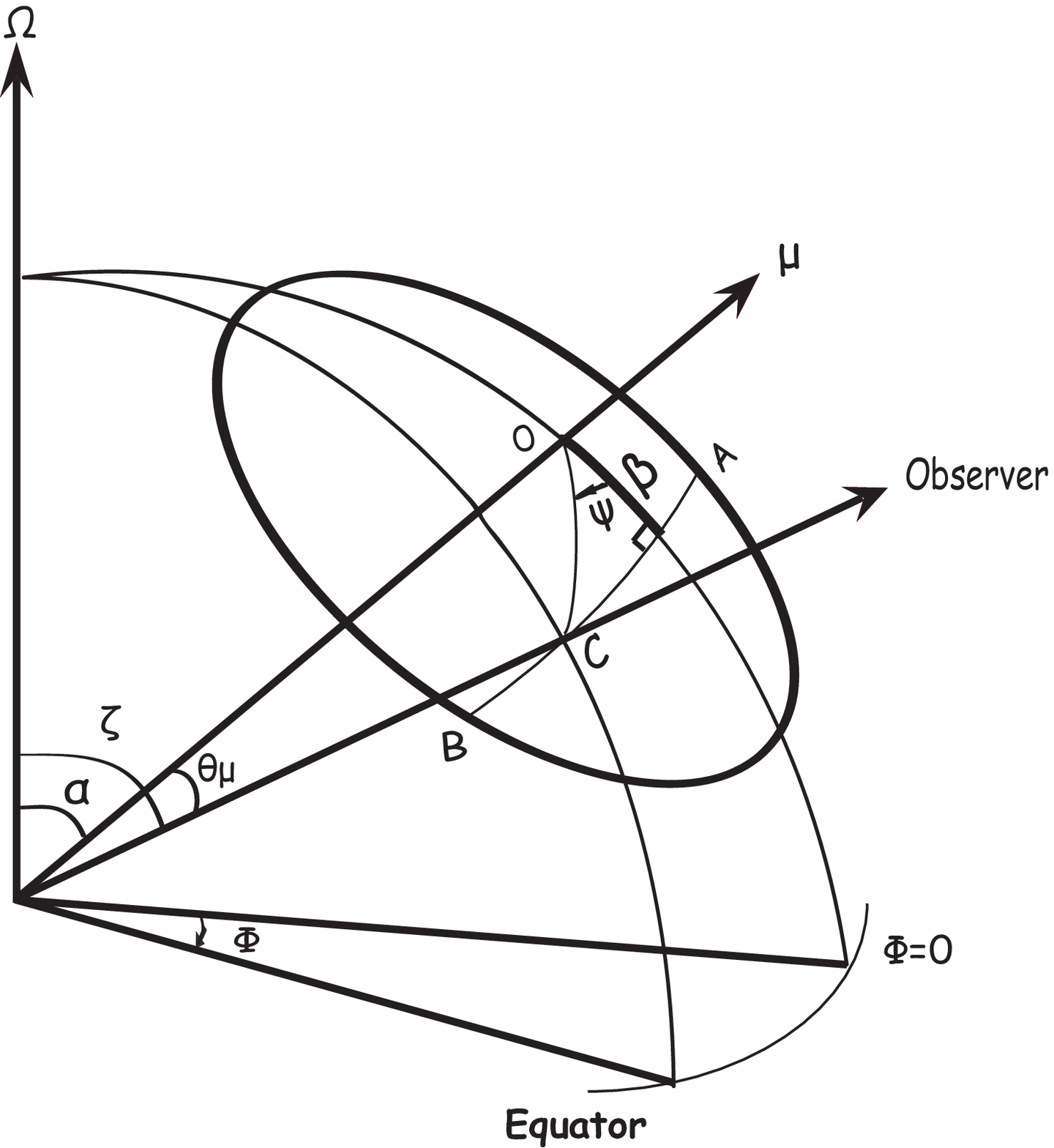}
\caption{The geometry of pulsar radiation beam. The line of sight sweeps the beam from A to B and the radiation is observed at C (Lyne \& Manchester 1988). \label{fig3}}
\end{figure}

\begin{figure}
\caption{The joint parameter space of the initial Lorentz factor $\gamma_{0}$ and the energy loss factor $\xi$. The grey areas represent the allowed parameter space derived at 95\% confidence level through fitting the index of $\theta_\mu-\nu$ relation, which is insensitive to inclination angles (LCW09). The solid-line and dotted-line contours represent the 2$\sigma$ confidence regions of $\gamma_0$ and $\xi$ constrained by fitting the absolute values of $\theta_\mu$ for different inclination angles, where the symbol ``+'' stands for the best fit $\gamma_0$ and $\xi$. For each pulsar, some inclination angles within the possible range given in literature  are selected as examples, their values are shown within the contour regions.  For a particular inclination angle, the joint parameter space is the overlapped region of grey area and the contour corresponding to the inclination angle. See Table 2 for details of the joint parameter space.  \label{fig4}}
\includegraphics[height=0.3\hsize]{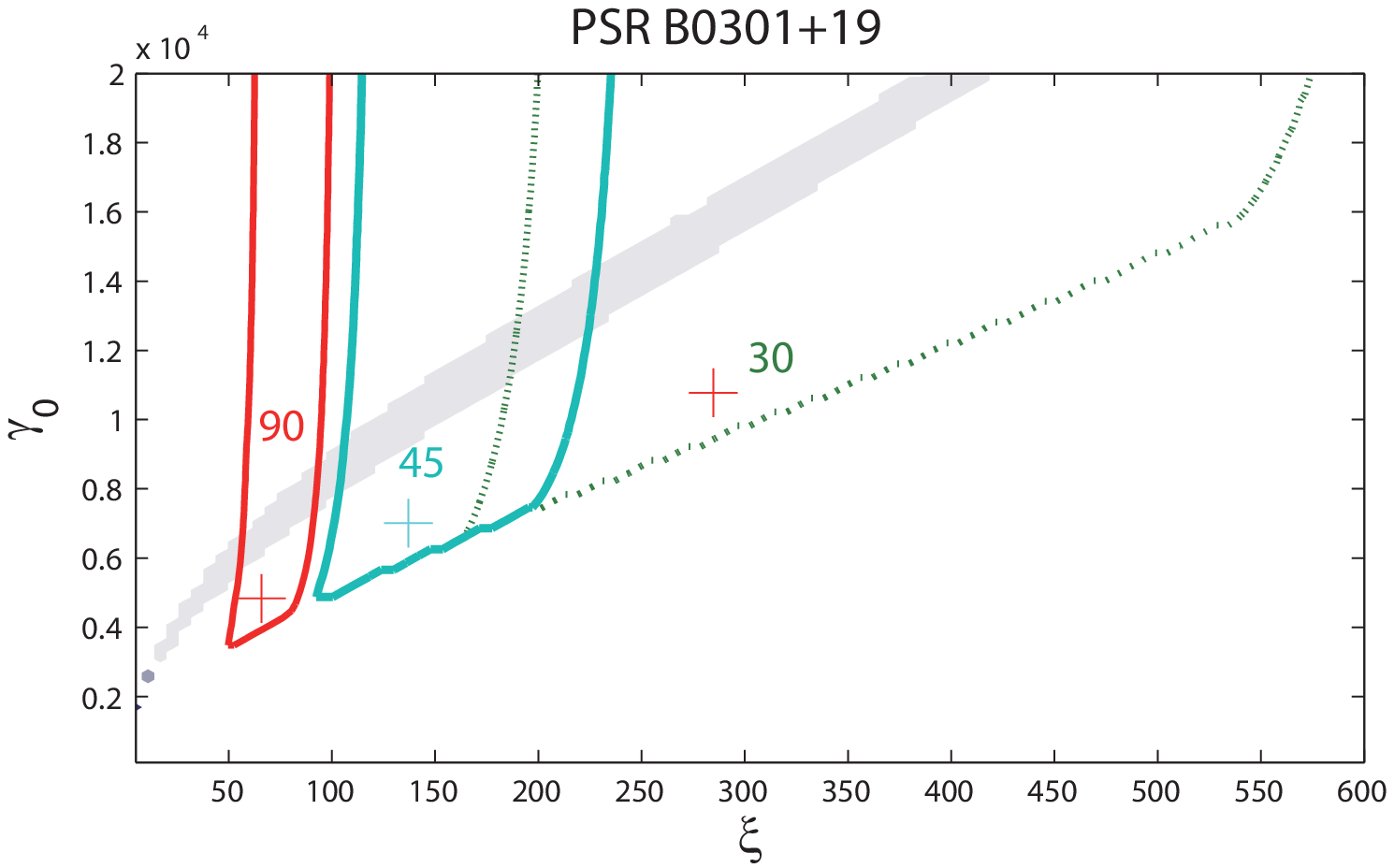}
\includegraphics[height=0.3\hsize]{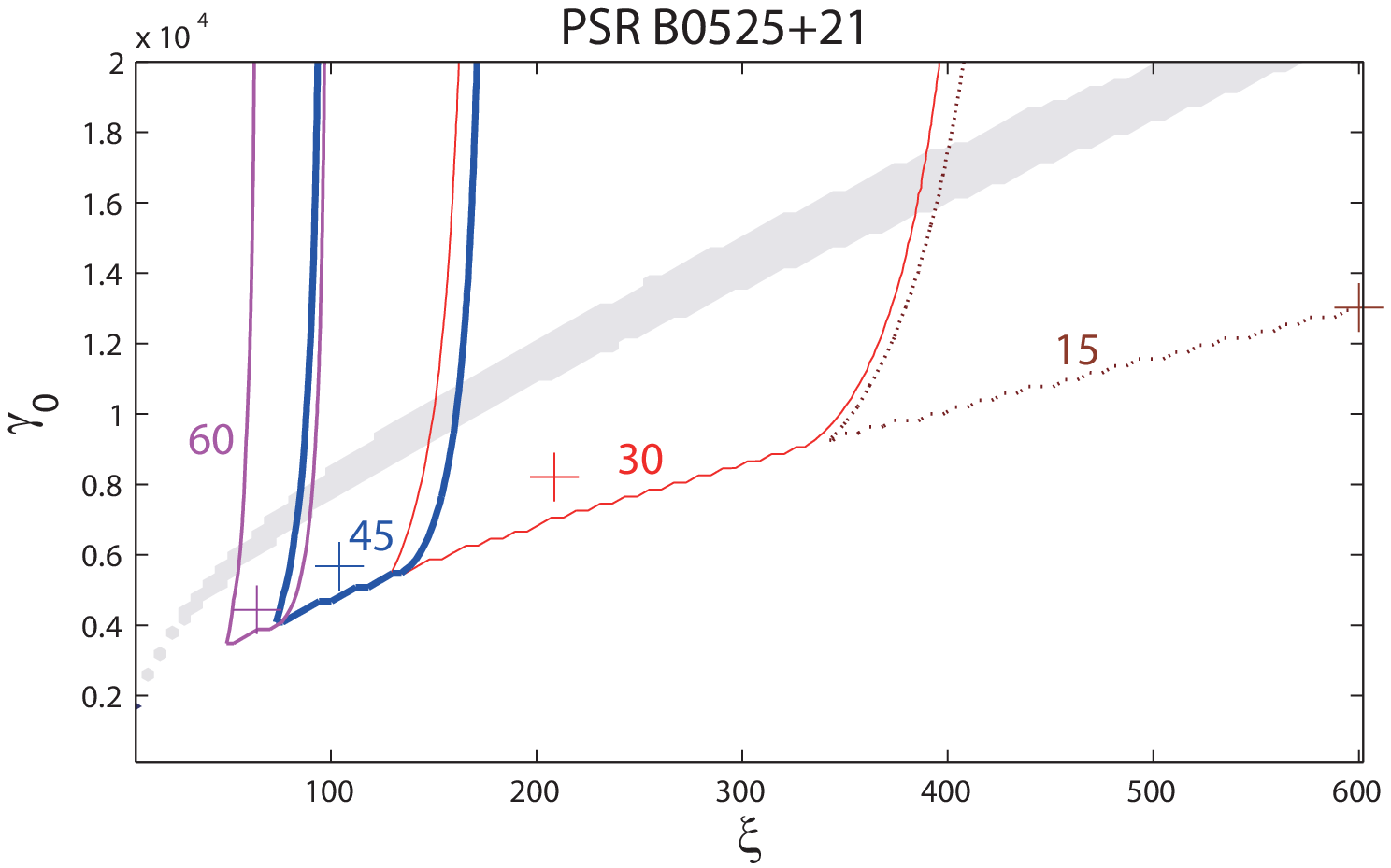}
\end{figure}
\clearpage

\begin{figure}
\includegraphics[height=0.3\hsize]{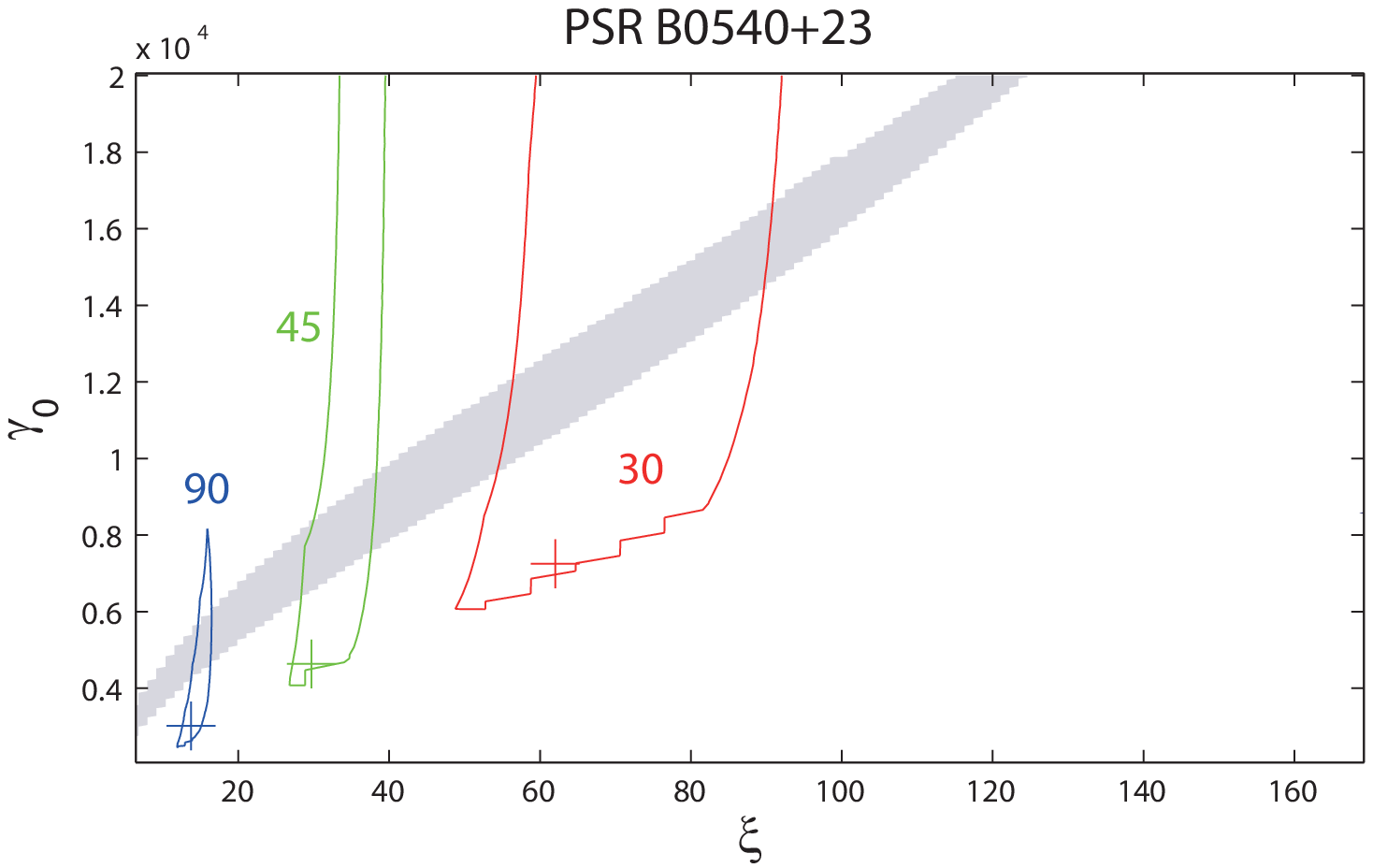}
\includegraphics[height=0.3\hsize]{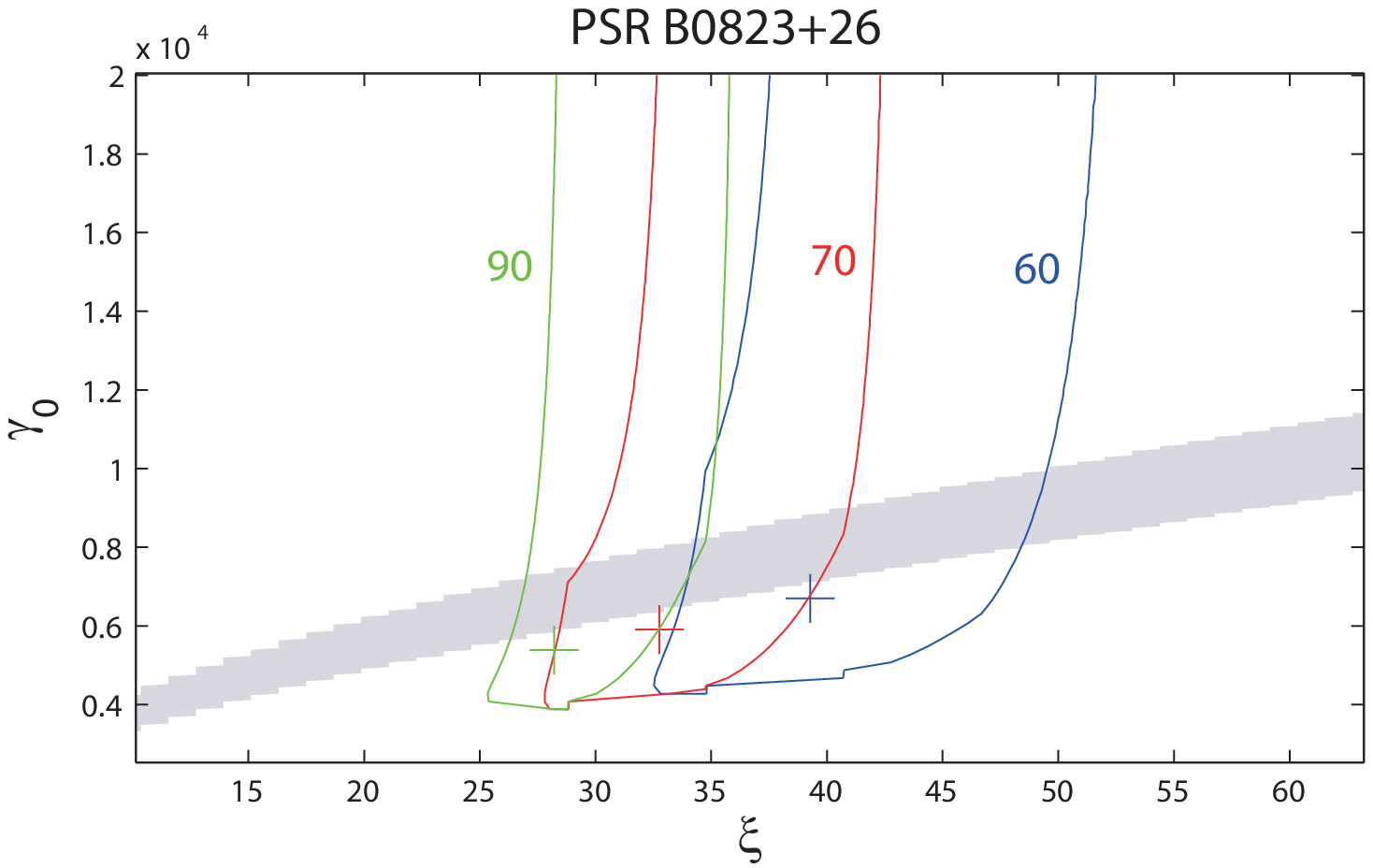}
\includegraphics[height=0.3\hsize]{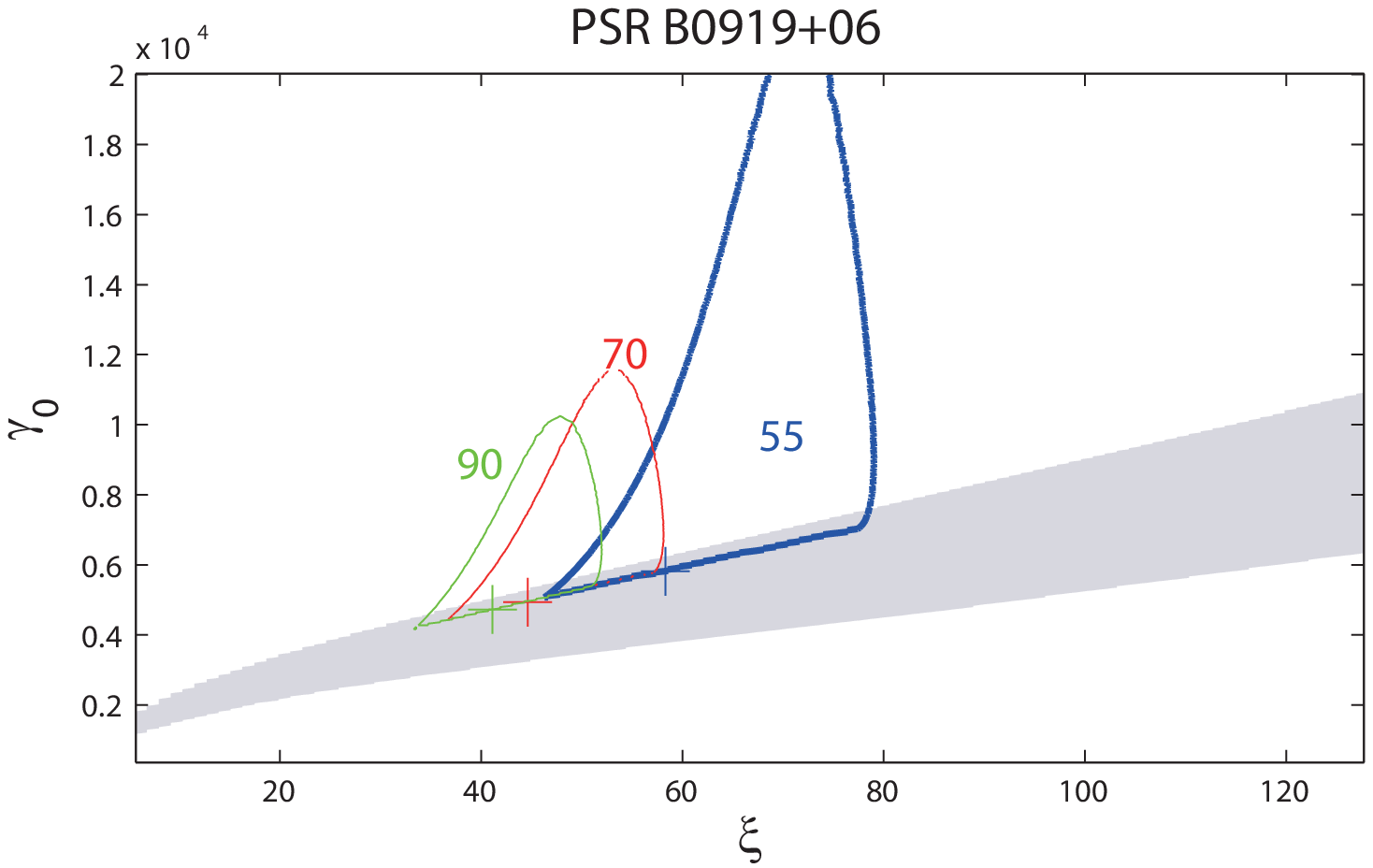}
\includegraphics[height=0.3\hsize]{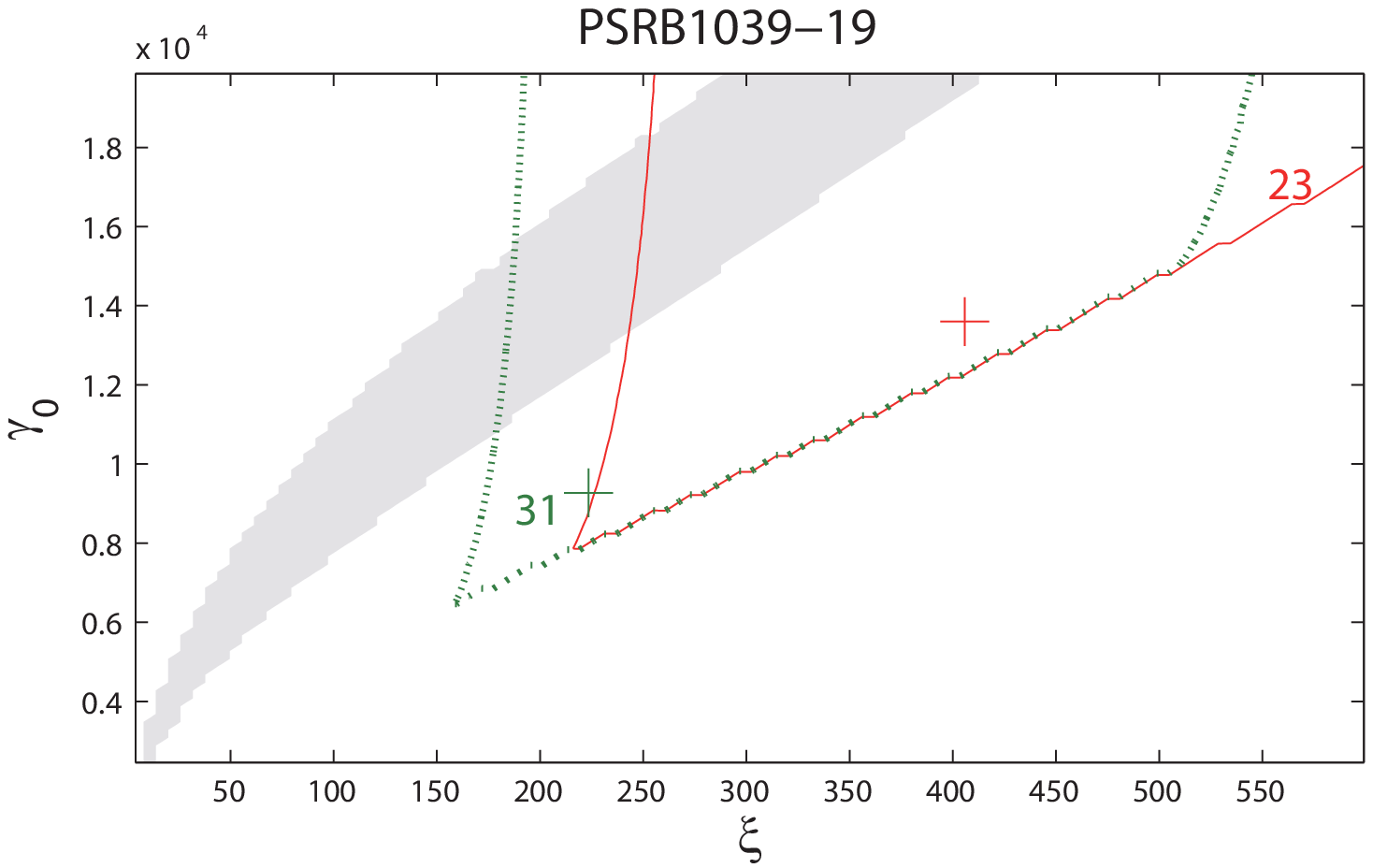}
\end{figure}
\clearpage

\begin{figure}
\includegraphics[height=0.3\hsize]{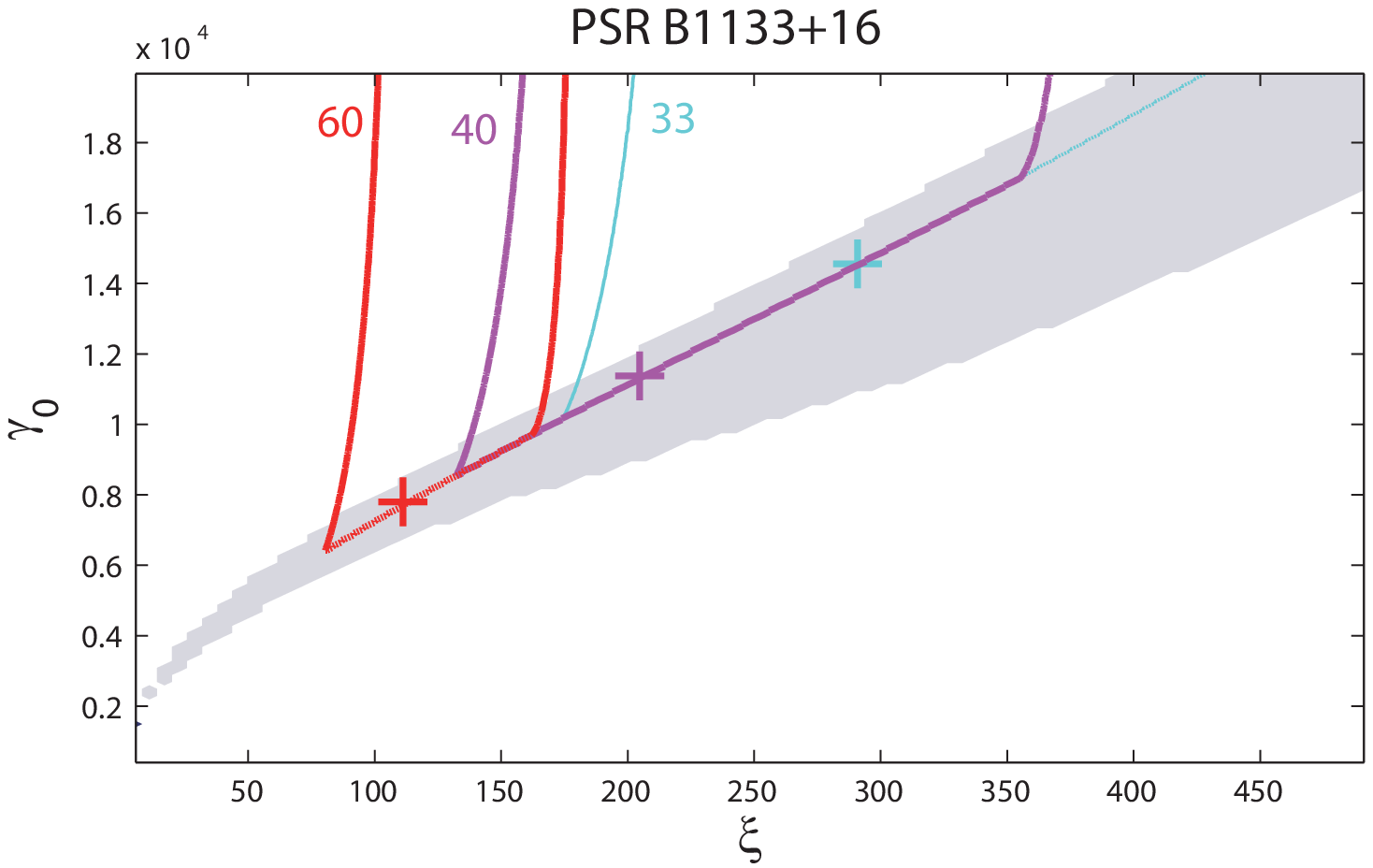}
\includegraphics[height=0.3\hsize]{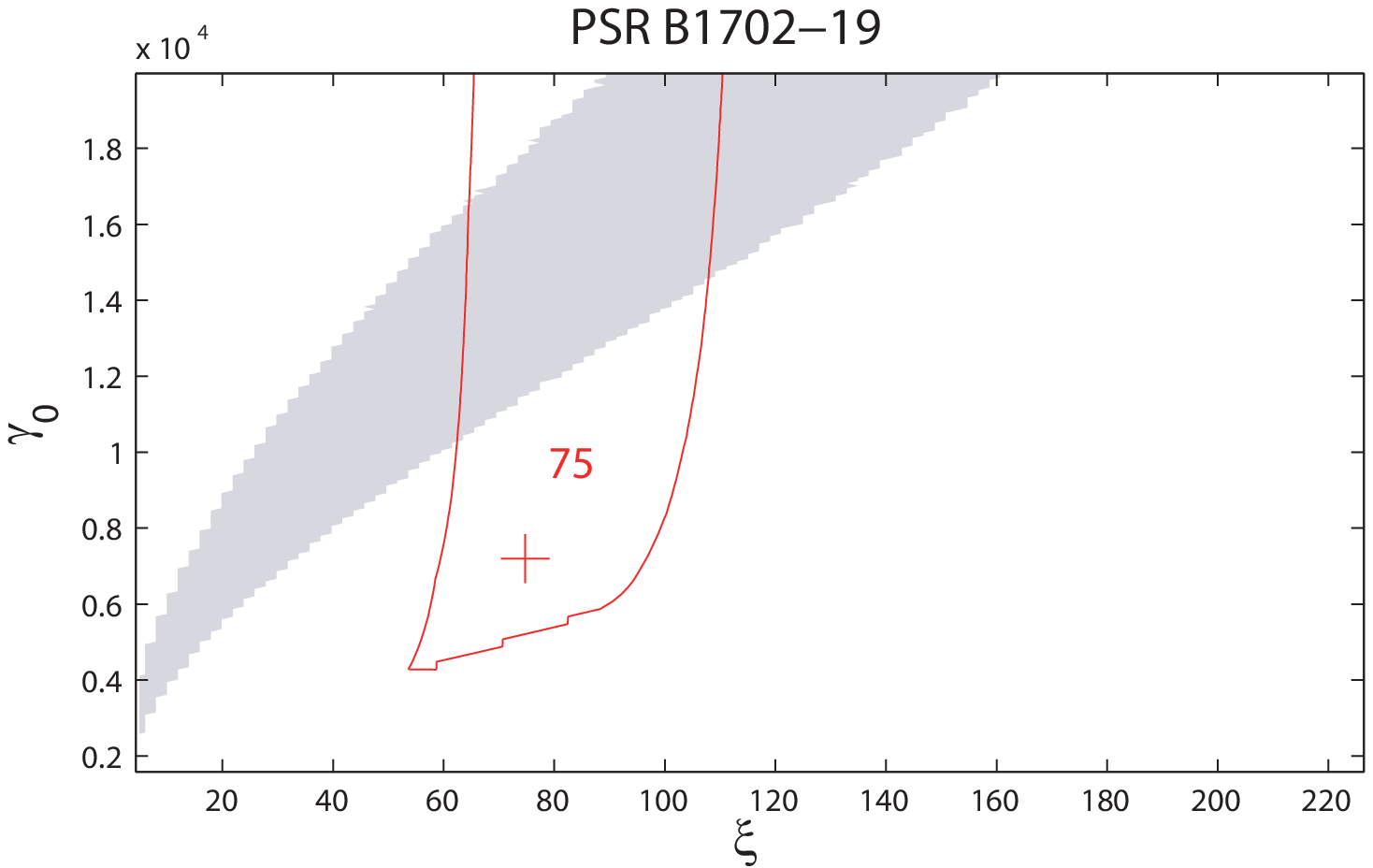}
\includegraphics[height=0.3\hsize]{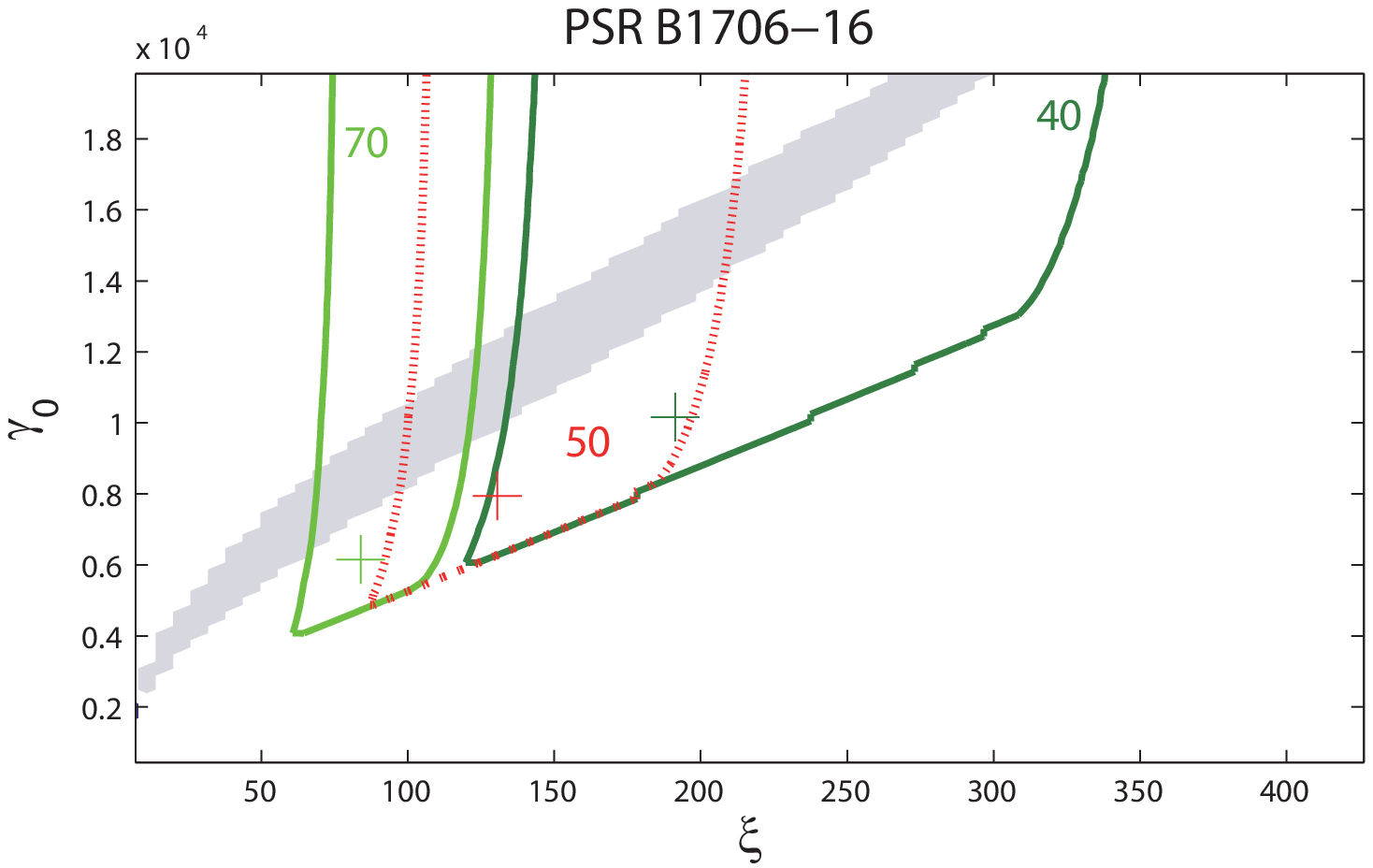}
\includegraphics[height=0.3\hsize]{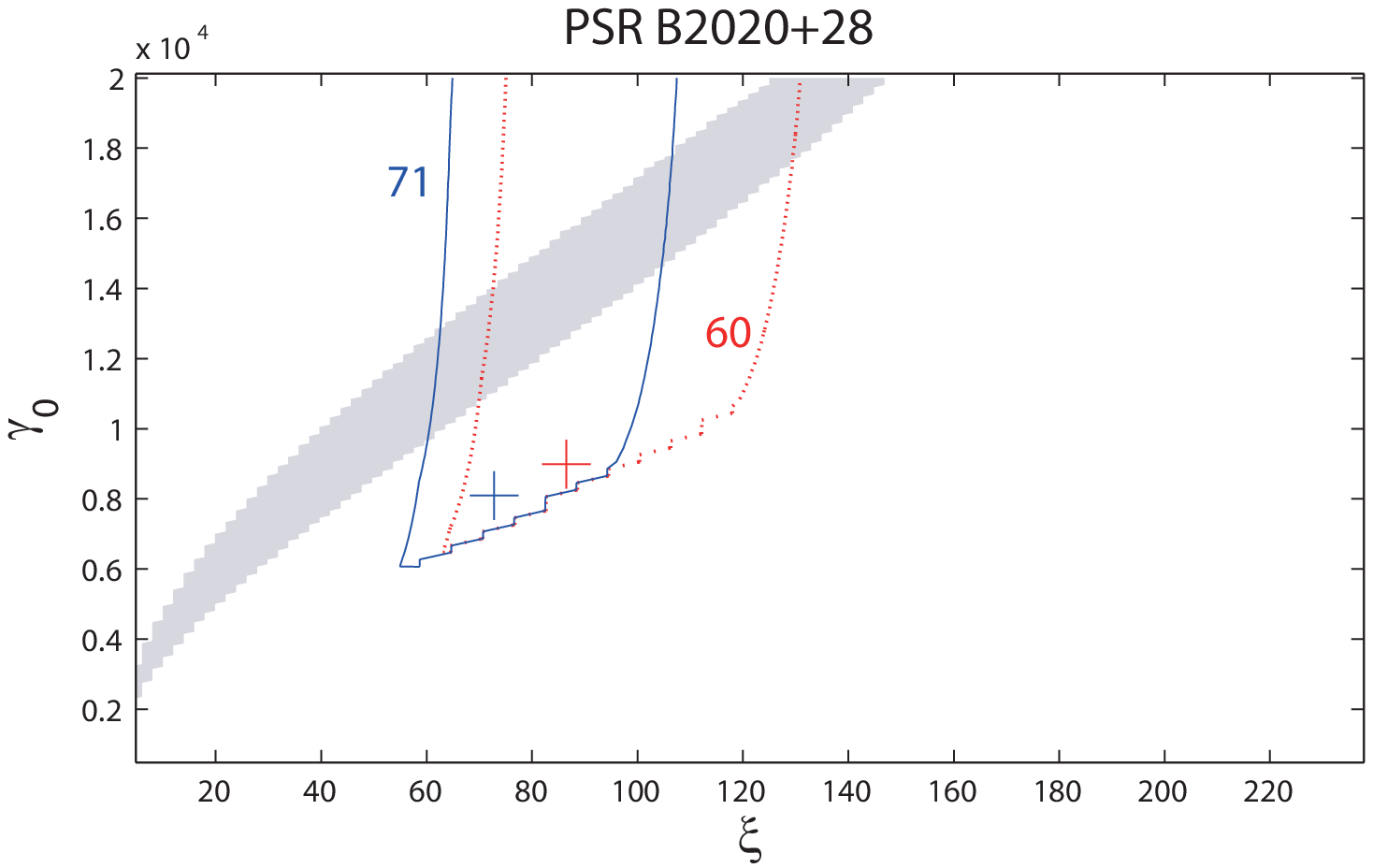}

\end{figure}
\clearpage

\begin{figure}
\includegraphics[height=0.3\hsize]{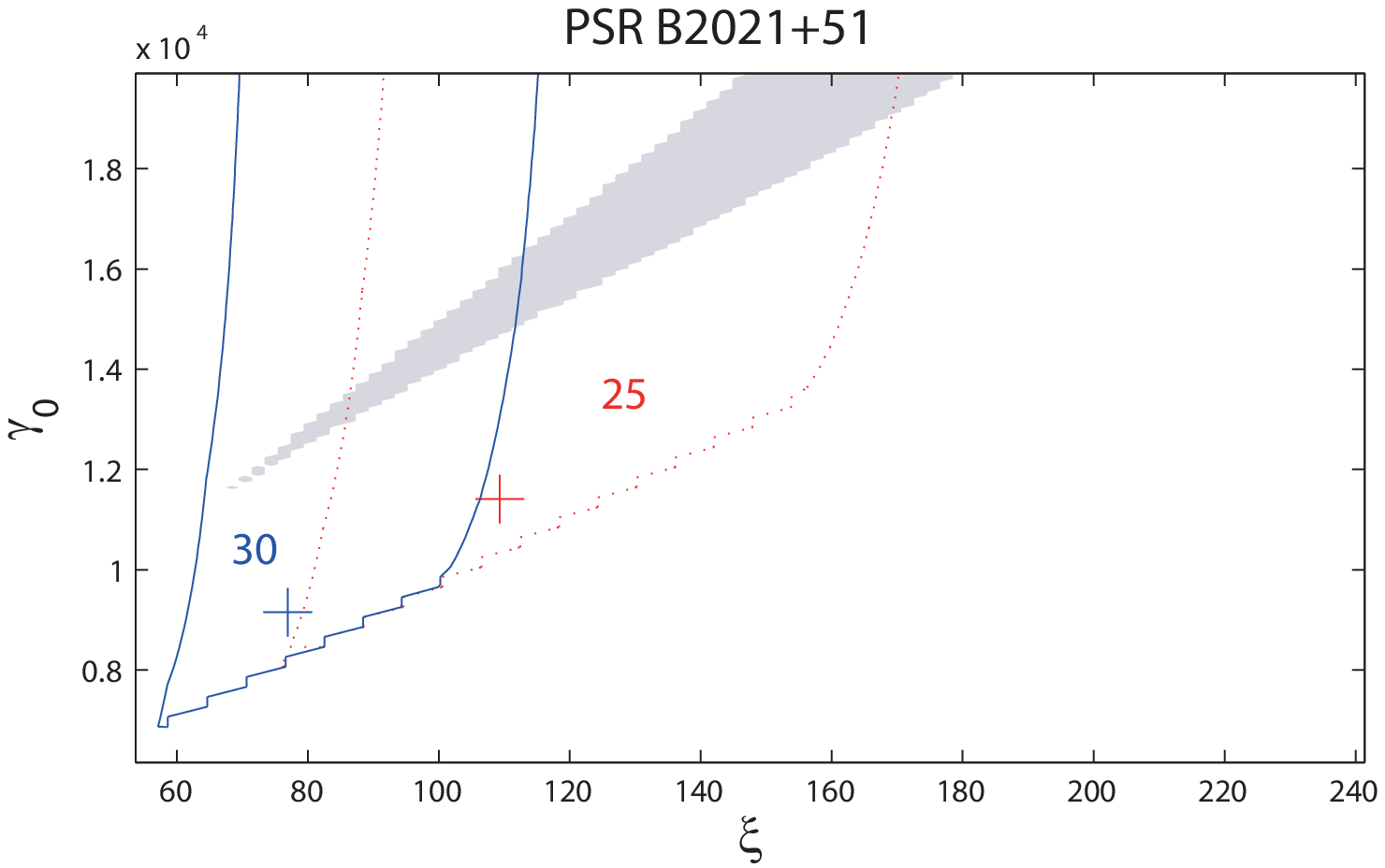}
\includegraphics[height=0.3\hsize]{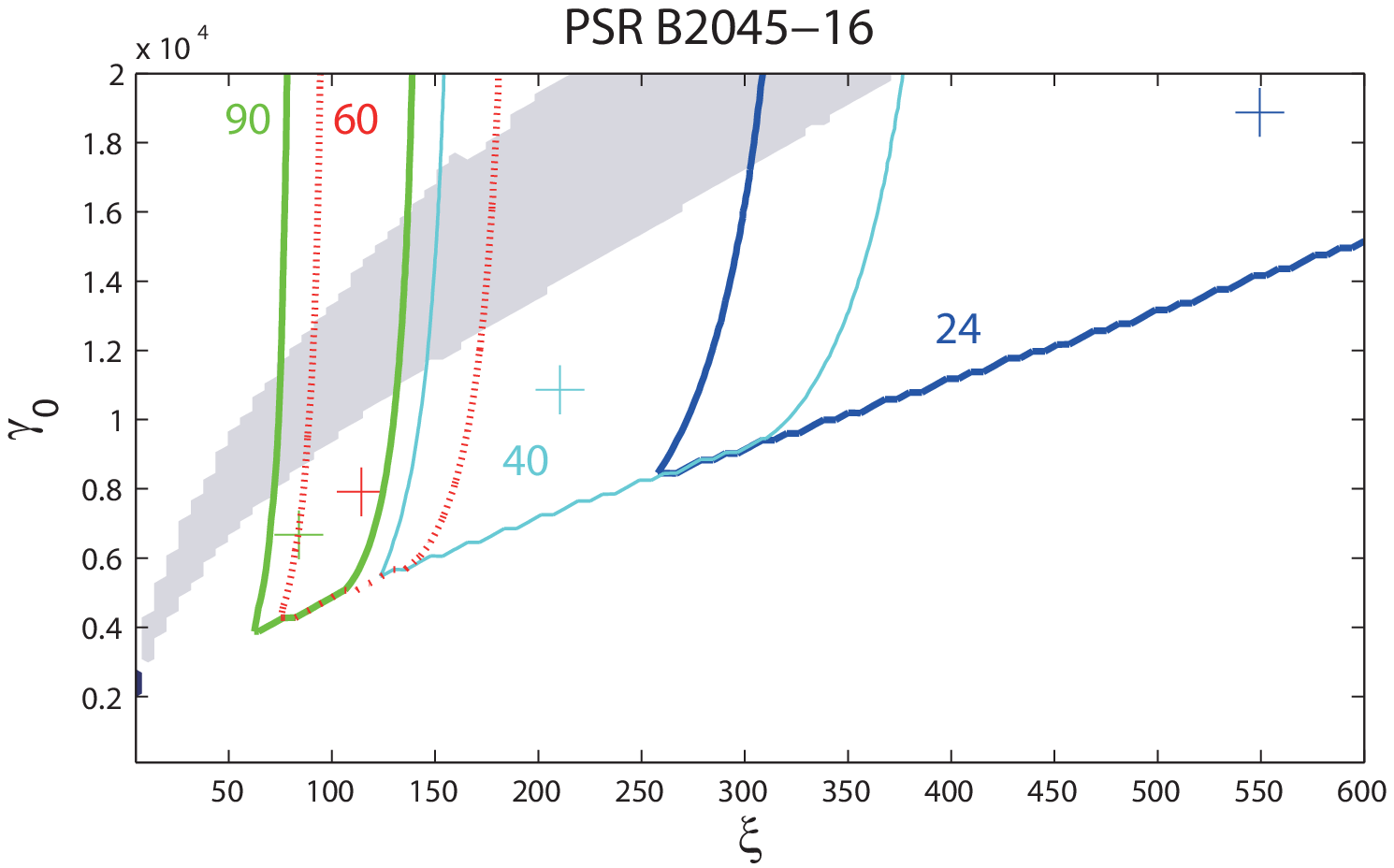}
\includegraphics[height=0.3\hsize]{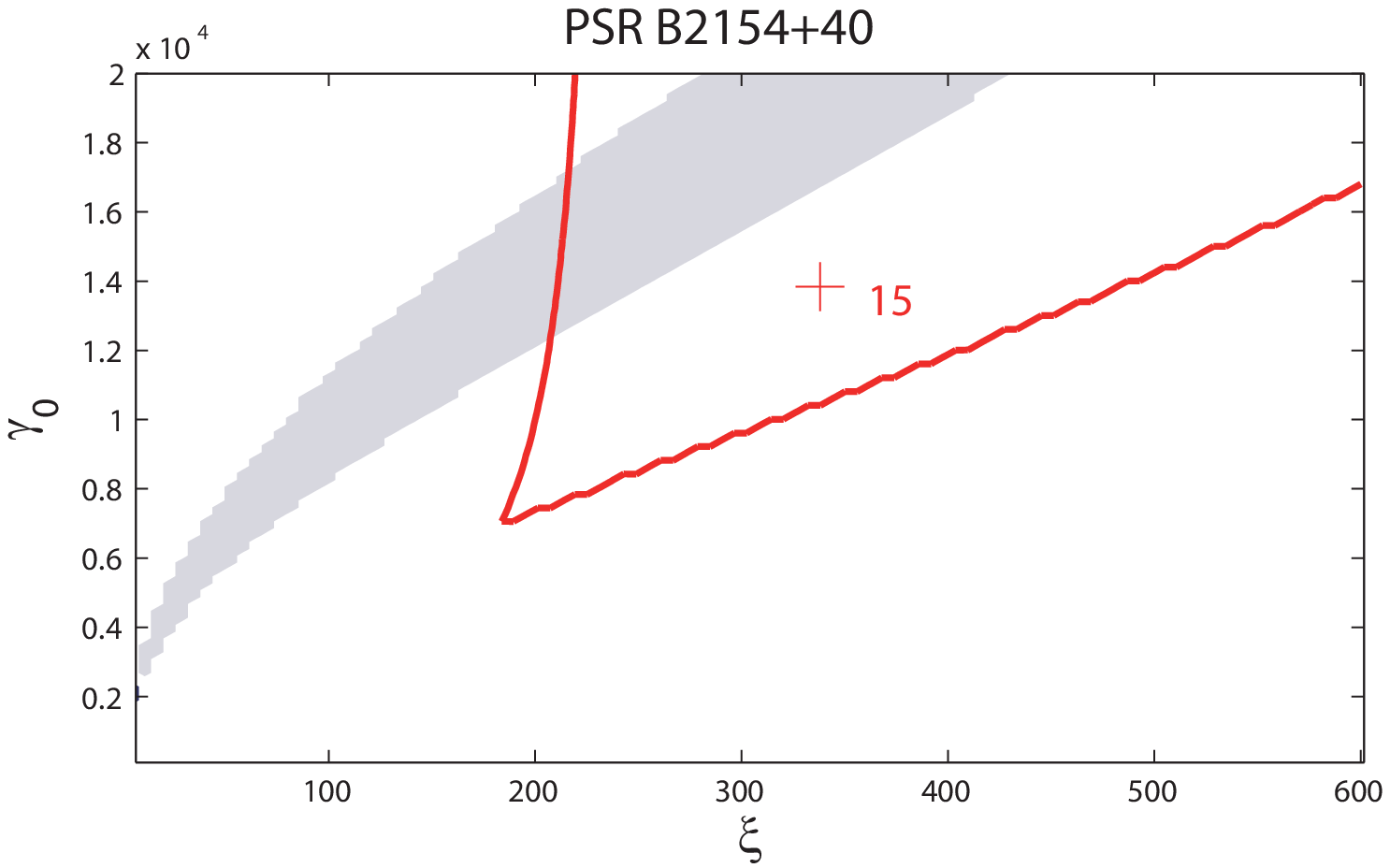}
\includegraphics[height=0.3\hsize]{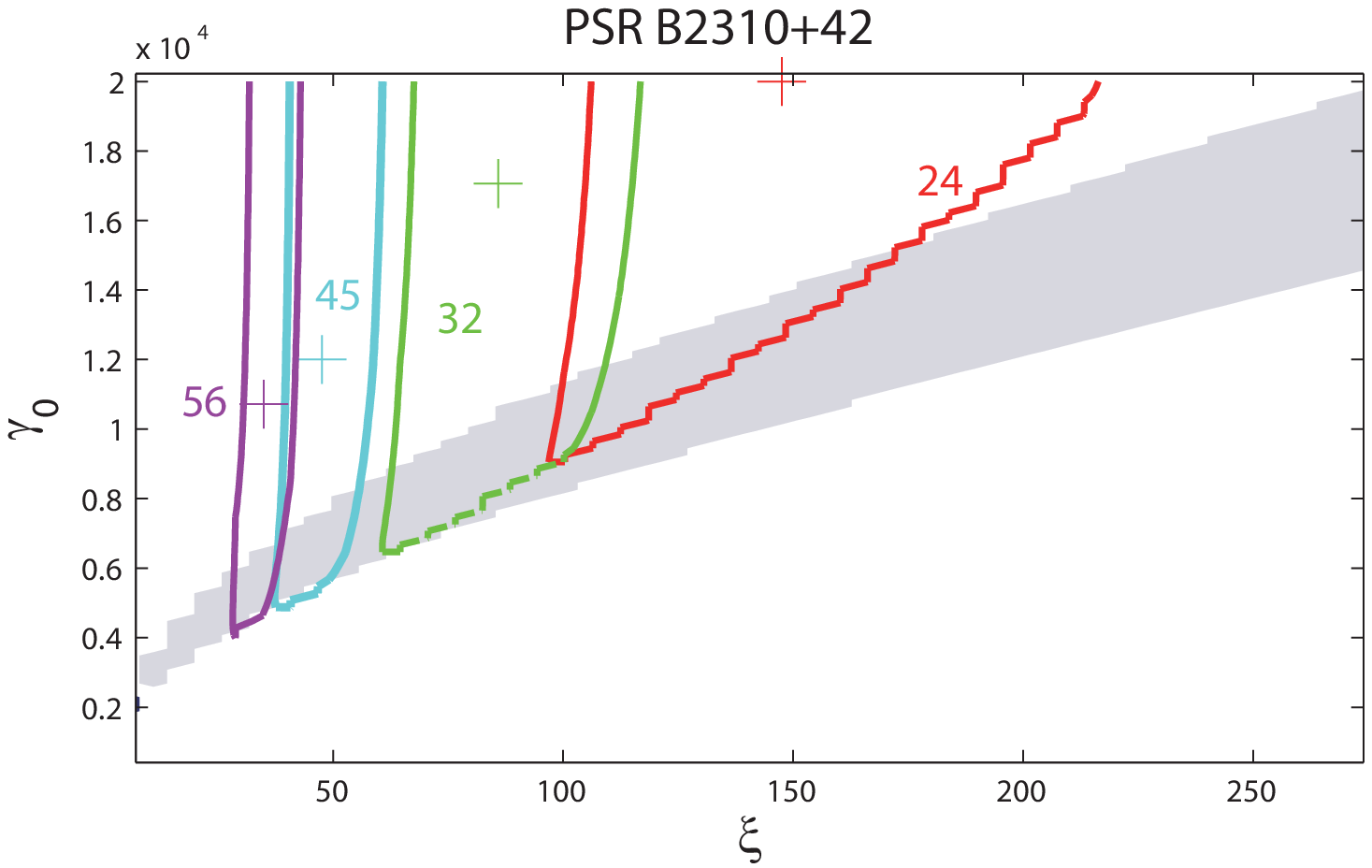}

\end{figure}
\clearpage

\begin{figure}
\includegraphics[height=0.3\hsize]{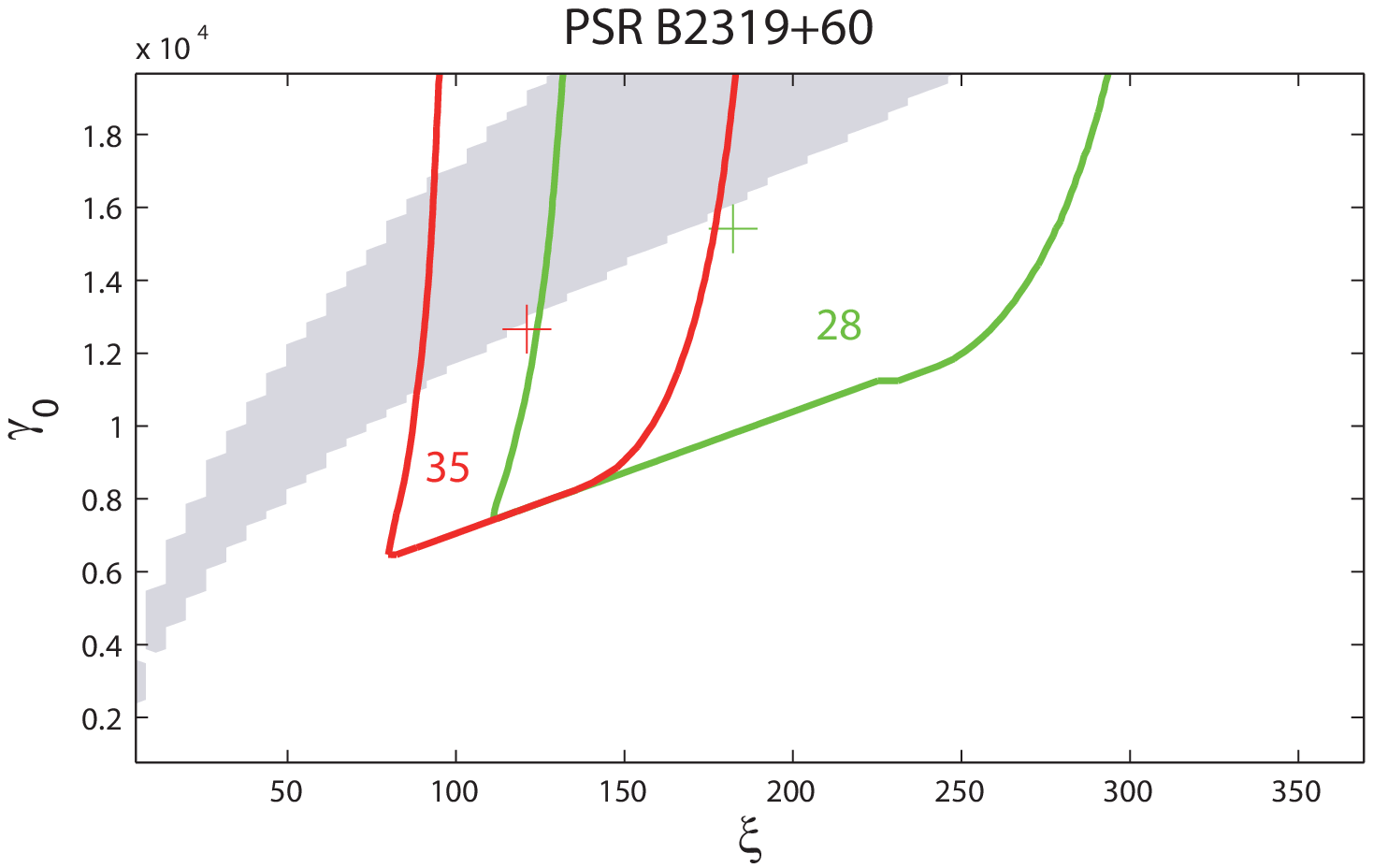}
\end{figure}

\end{document}